\documentclass[twocolumn,citeautoscript,pra,superscriptaddress,amsmath,amssymb]{revtex4-1}
 
\usepackage{graphicx}
\usepackage{color}
\usepackage{upgreek}
\usepackage{float}
\usepackage{lipsum}
\usepackage{mathrsfs}
\usepackage{bm}
\usepackage[normalem]{ulem}
\usepackage[colorlinks,citecolor=blue,linkcolor=blue]{hyperref}
 
\newcommand{\ket}[1]{\ensuremath{\left| #1 \right\rangle}}
\newcommand{\bra}[1]{\ensuremath{\left\langle #1 \right|}}
\newcommand{\kket}[1]{\ensuremath{\left| #1 \right\rangle\!\rangle}}
\newcommand{\bbra}[1]{\ensuremath{\left\langle\!\langle #1 \right|}}
\DeclareMathOperator*{\argmax}{\arg\!\max}
\DeclareMathOperator*{\argmin}{\arg\!\min}

\begin{document}
 
\title{Demonstration of non-Markovian process characterisation \\ and control on a quantum processor}
 
\author{G. A. L. White}
\email{white.g@unimelb.edu.au}
\affiliation{School of Physics, University of Melbourne, Parkville, VIC 3010, Australia}

\author{C. D. Hill}
\email{cdhill@unimelb.edu.au}
\affiliation{School of Physics, University of Melbourne, Parkville, VIC 3010, Australia}
\affiliation{School of Mathematics and Statistics, University of Melbourne, Parkville, VIC, 3010, Australia}

\author{F. A. Pollock}
\email{felix.pollock@monash.edu}
\affiliation{School of Physics and Astronomy, Monash University, Clayton, VIC 3800, Australia}

\author{L. C. L. Hollenberg}
\email{lloydch@unimelb.edu.au}
\affiliation{School of Physics, University of Melbourne, Parkville, VIC 3010, Australia}

\author{K. Modi}
\email{kavan.modi@monash.edu}
\affiliation{School of Physics and Astronomy, Monash University, Clayton, VIC 3800, Australia}

\begin{abstract}
\section*{Abstract}
In the scale-up of quantum computers, the framework underpinning fault-tolerance generally relies on the strong assumption that environmental noise affecting qubit logic is uncorrelated (Markovian). However, as physical devices progress well into the complex multi-qubit regime, attention is turning to understanding the appearance and mitigation of correlated — or non-Markovian — noise, which poses a serious challenge to the progression of quantum technology. This error type has previously remained elusive to characterisation techniques. Here, we develop a framework for characterising non-Markovian dynamics in quantum systems and experimentally test it on multi-qubit superconducting quantum devices. Where noisy processes cannot be accounted for using standard Markovian techniques, our reconstruction predicts the behaviour of the devices with an infidelity of $10^{-3}$. Our results show this characterisation technique leads to superior quantum control and extension of coherence time by effective decoupling from the non-Markovian environment. This framework, validated by our results, is applicable to any controlled quantum device and offers a significant step towards optimal device operation and noise reduction.
\end{abstract}
 
\maketitle
 
\section*{Introduction}
The theoretical machinery for open quantum system dynamics is well-oiled in low-coupling cases, but strong environmental interactions can lead to non-trivial dynamical memory effects that are difficult to understand, much less control. 
The recent advent of high performance quantum information processors~(QIPs) has precipitated greater sensitivity to  complex dynamical effects.
In particular, it is clear that device behaviour must be understood under a relaxed Markov assumption~\cite{Haase2018, IBMStateFid, Sarovar-crosstalk}. 
The resulting non-Markovian dynamics includes more general errors that may be temporally correlated or dependent on broader environmental context~\cite{Rivas2014, Li2018, fattah}.
Characterisation techniques of quantum devices such as randomised benchmarking~(RB) and gate set tomography~(GST) have so far represented the front line in understanding and addressing noise~\cite{PhysRevLett.78.390, gst-2013, sandia-GST, PhysRevA.87.062119, White2019}.
However, constructing a digestible picture of non-Markovian behaviour has proven difficult, and violates the error model assumed in these methods.
Chiefly, this is because quantum correlations can forbid the division of dynamical processes into arbitrary steps of completely positive (CP), linear maps~\cite{Pechukas1994}.
If information back-flow from the environment can occur, then noisy effects can be influenced by past factors; this detail can no longer be `forgotten'.
\par 
For device control, this is problematic.
The circuit model of quantum computation is predicated on identical gates implemented at different times having identical actions.
Markovian errors multiply out and propagate in predictable ways.
However, non-Markovian noise gives rise to adverse effects that are much more challenging to tame. 
For example, correlated errors can spread across the device, and have been shown to lower thresholds of quantum error correcting codes~\cite{correlated-qec,PhysRevA.99.052351}.
Similarly, context-dependent gates allow for poorly understood forms of dynamical errors not describable by a Markov model. 
This is one of the largest obstacles to near-term QIPs; non-Markovian noise must be either eliminated or, as some have suggested, harnessed into a resource~\cite{non-Markovian-control,Pineda2016,Kumar2018,Bylicka2013,Berk2019}.
Until recently, there has not been a clear operational definition for quantum non-Markovianity, nor consensus that one unifying measure could even be found. 
\par
Using the recent process tensor framework~\cite{Pollock2018a}, we develop a robust device characterisation technique which is inclusive of non-Markovian dynamics.
We keep discussion fully general, but demonstrate the capabilities of this method on four different IBM Quantum superconducting quantum devices. 
We then examine the robustness of the framework's assumptions; address shortcomings; and demonstrate its functionality in process characterisation, memory detection, and application to adaptive quantum control.
We find that we can characterise arbitrary processes down to an average infidelity of $10^{-3}$-- quantifying its predictive power for the future states of the system given some past operations.
We show that this outperforms the characterisation given by the standard technique of GST in the presence of non-Markovian effects, which employs a comprehensive Markov model.
With non-Markovian dynamics fully accounted for, we discuss applications of the process tensor generically to adaptive quantum control. 
As an example, we demonstrate how two qubits can be decoupled without any \emph{a priori} knowledge or assumption about their interactions, and how typically inaccessible user-designated non-unitary control operations can be realised.
The efficacy of this framework over a range of devices showcases its consistency and broad-range of applicability.
Our results represent significant progress towards the characterisation and optimal control of non-Markovian QIPs and other quantum devices.
\par

\begin{figure*}
    \centering
    \includegraphics[width=\linewidth]{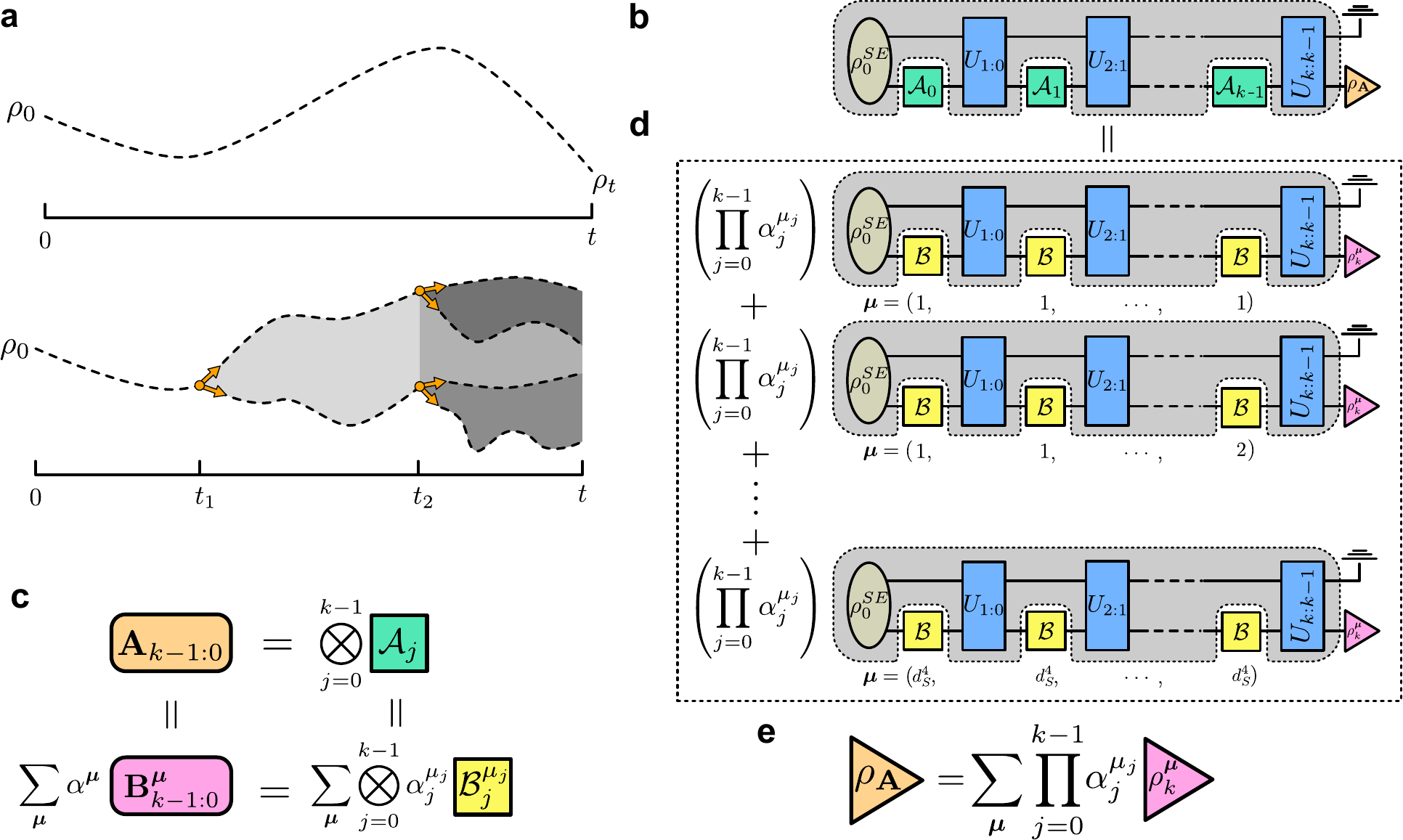}
    \caption{\textbf{An illustrative summary of process characterisation.} \textbf{a} The state of an open system over time follows a trajectory through state space until some final time at which the state is probed (top). By applying control operations at times $t_1$ and $t_2$, an experimenter can anchor and change the trajectory, which can be inferred via a linear combination of trajectories corresponding to basis operations (bottom).
    \textbf{b} A circuit model showing a sequence of operations $\{\mathcal{A}_j\}$ interleaved with $SE$ interactions, resulting in a final state $\rho_{\mathbf{A}}$.
    \textbf{c}
    A sequence of operations $\mathbf{A}_{k-1:0}$ can be expressed as a tensor product of independently chosen operations $\mathcal{A}_j$ at each time step. These can then be individually decomposed into a chosen basis $\{\mathcal{B}_j^{\mu_j}\}$ together giving a basis of sequences $\{\mathbf{B}_{k-1:0}^{\bm{\mu}}\}$.
    \textbf{d}
    A process can be fully characterised by measuring the output state for a complete set of basis operations at different times. Then, an arbitrary process can be expressed as a linear combination of each basis process; because of the linear construction, the intermediate evolution is completely preserved in the description of the arbitrary process. 
    \textbf{e} The final state density matrix for the process $\textbf{A}_{k-1:0}$ can be expressed by tracing over all of the intermediate operations, contracting to a coefficient expansion for the measured density matrices in the basis processes. This is the same density matrix as in \textbf{b}.}
    \label{fig:PT_summary}
\end{figure*}
\section*{Results}
\subsection*{Process Characterisation}
\label{ssec:NM-processes}
To characterise non-Markovian device features, we employ the process tensor framework, which was recently developed to describe arbitrary quantum processes.
Non-Markovian dynamics constitute any interaction between a system and its environment which then affects the system at a later time; the environment need not even be coherent. For superconducting processors, this behaviour for example could stem from coupling with neighbouring qubits, leakage into higher energy levels, or two-level-system defects \cite{Muller2019}.
Here, we briefly outline some relevant background before detailing our approach to the problem.
Traditional approaches to quantum stochastic dynamics are concerned with tracking the state of the system~($S$) as a function of time: $\rho_t = \text{tr}_E[U_{t:0}\, (\rho^{SE}_0) ]$, where $U(\cdot) = u (\cdot) u^\dag$ is a unitary map on system-environment ($SE$), initially in state $\rho^{SE}_0$ (often required to be uncorrelated). However, real experiments are driven by  sequences of control operations, mathematically represented by trace non-increasing CP maps $\{\mathcal{A}_0, \dots, \mathcal{A}_{k-1}\}=: \textbf{A}_{k-1:0}$. The process tensor is designed to account for the intermediate control operations and quantifies quantum correlations between past events and the final state of the system. 
In doing so, the process tensor formally generalises the notion of a stochastic process to the quantum domain~\cite{Milz2020kolmogorovextension} and reduces to a classical stochastic process in the correct limit~\cite{arXiv:1905.03018, arXiv:1907.05807}. 
The formalism gives rise to a clear necessary and sufficient definition of quantum non-Markovianity~\cite{Pollock2018}, as well as other features of non-Markovian memory~\cite{taranto1, taranto2, taranto3}.
Figure~\ref{fig:PT_summary}a, top and bottom, illustrates respectively the traditional approach and the process tensor approach to describing a quantum process. 
In the top panel, a quantum state left to evolve in isolation can be reconstructed at $t$ via quantum state tomography~(QST).
In the bottom panel, events come in the form of control operations applied to $S$ at times $t_1$ and $t_2$; the future states of the $S$ branch at time $t$ are conditioned on the outcomes of the control operations.
\par
Mathematically, the controlled dynamics has the form
\begin{gather}
    \rho_k\left(\textbf{A}_{k-1:0}\right) = \text{tr}_E [U_{k:k-1} \, \mathcal{A}_{k-1} \cdots \, U_{1:0} \, \mathcal{A}_{0} (\rho^{SE}_0)],
\end{gather}which can be rearranged, as depicted in Figure~\ref{fig:PT_summary}b, to define a mapping from past control to future states: $\rho_k\left(\textbf{A}_{k-1:0}\right) = \mathcal{T}^{k:0}[\textbf{A}_{k-1:0}]$. The process tensor, $\mathcal{T}^{k:0}$ is a multi-linear map on the control operations, and includes all of the information hidden to the experimenter, including correlations in the initial state, and any intermediate interaction with the environment.
\par
The set of possible sequences of CP maps $\mathbf{A}_{k-1:0}$ forms a product vector space, built up from the spaces of temporally local operations; in particular, $\mathbf{A}_{k-1:0} = \bigotimes_{j=0}^{k-1} \mathcal{A}_j$ when the operations at each time are chosen independently. 
As such, the process tensor is completely characterised by its input-output relations on a complete basis of control operations, just as a quantum channel is unambiguously defined by its input-output relations on a complete basis of states. 
Let us denote the basis for CP maps at the $j$th time step as $\{\mathcal{B}_j^{\mu_j}\}_{\mu_j=1}^{d_S^4}$ and the basis sequences as $\{\mathbf{B}_{k-1:0}^{\bm{\mu}}\}_{\bm{\mu}=(1,1,\cdots,1)}^{(d_S^4,d_S^4,\cdots,d_S^4)}$ such that an arbitrary sequence of operations can be written as $\mathbf{A}_{k-1:0} = \sum_{\bm{\mu}} \alpha^{\bm{\mu}} \ \mathbf{B}_{k-1:0}^{\bm{\mu}}$, see Figure~\ref{fig:PT_summary}c. 
Then the process tensor's action is defined by
\begin{gather}
\label{DM-expansion}
\rho_k(\mathbf{A}_{k-1:0}) \!=\! \sum_{\bm{\mu}} \alpha^{\bm{\mu}} \ \rho_k^{\bm{\mu}} \quad \text{with} \quad
\rho_k^{\bm{\mu}} \!:=\!
\mathcal{T}^{k:0}[\textbf{B}^{\bm{\mu}}_{k-1:0}].
\end{gather}
In other words, to reconstruct the process tensor we need to experimentally estimate $\rho_k^{\bm{\mu}}$ for all $\bm{\mu}$, this is depicted in Figure~\ref{fig:PT_summary}d. 
A key assumption to this model is that the relationship between the gates acting on the system is ideal, and that the duration of the gates is small when compared to the overall dynamics of the system. For superconducting devices, single qubit gates are short, with fidelities of $\mathcal{O}(10^{-4})$, and so we do not expect this to be a problem. This assumption will need to be revisited for the case of two-qubit gates, however.
In the methods section, we detail explicitly the steps to go from $\rho_k^{\bm{\mu}}$ to constructing the process tensor.
Once the process tensor is reconstructed, using Equation~\eqref{DM-expansion}, one can predict the final density matrix corresponding to any choice of control sequence $\mathbf{A}_{k-1:0}$, as shown in Figure~\ref{fig:PT_summary}e.
We use prediction fidelity of the final states, conditioned on controls, as a performance metric for our process characterisation.
\subsection*{Experimental Implementation}
\label{sec:process-characterisation}
We look now to the practical determination of the process tensor in experiment. The experiments carried out in this work used cloud-based IBM Quantum superconducting quantum devices. We first evaluated predictive capabilities of process tensor over a host of different experiments on the IBM Quantum devices ibmq\_johannesburg (shortened: `Jo'burg'), ibmq\_boeblingen (`Boeb.'), ibmq\_poughkeepsie (`PK'), and ibmq\_valencia. Our main contribution is in demonstrating how this framework leads to high fidelity process characterisation and precise control over non-Markovian dynamics. 
Ideally, complete process tensor construction would be achieved with the full span of CP maps.
Unfortunately, efficient measurement within the coherence time is beyond the scope of most current hardware. 
For now, this rules out non-unital and trace-decreasing maps on superconducting devices, affording only unitary control, i.e., we do not have a complete basis of operations. With these limitations, processes can still be characterised in terms of `restricted' process tensors $\mathcal{T}^{k:0}_{\text{r}}$~\cite{PT-limited-control}, defined in a similar way to the full process tensor, but constrained to act only on the subspace of operations comprising the linear span of unitary maps. This reduces the control space to $d_S^4 - 2d_S^2 + 2$ dimensions. 
We reconstruct and test the four-time restricted process tensor $\mathcal{T}^{3:0}_r$ for a single qubit process on IBM Quantum devices. To do so, we first reconstruct the final quantum states $\rho^{ijk}_3$. This state depends on the past controls, i.e., the initial preparation $\mathcal{P}^i_0 \in \mathcal{P}$ and the subsequent unitaries $\mathcal{U}^j_1 \in \mathcal{U}$ and $\mathcal{U}^k_2 \in \mathcal{U}$. The restricted process tensor is then obtained using Equation \eqref{DM-expansion}. 
The set $\mathcal{U}$ contains 28 random unitaries, where the first $n$ elements $\mathcal{U}^{(n)}$ are used to reconstruct $\mathcal{T}^{3:0}_r$. Each smaller basis $\mathcal{U}^{(n)}$ is a subset of the larger bases. Randomly chosen unitaries are almost surely linearly independent, and are selected so as not to systematically preference any part of superoperator space.
The remaining $28-n$ elements are contracted with the reconstructed $\mathcal{T}^{3:0}_r$ to obtain predictions $\sigma_3^{ijk}$. 
We then compute the reconstruction fidelity
\begin{gather}
\mathcal{F}_{ijk} := \left[\text{tr}\sqrt{\sqrt{\rho_3^{ijk}}\sigma_3^{ijk}\sqrt{\rho_3^{ijk}}}\right]^2
\end{gather}
to gauge the accuracy of the prediction.\par
\begin{figure}
    \centering
    \includegraphics[width=0.95\linewidth]{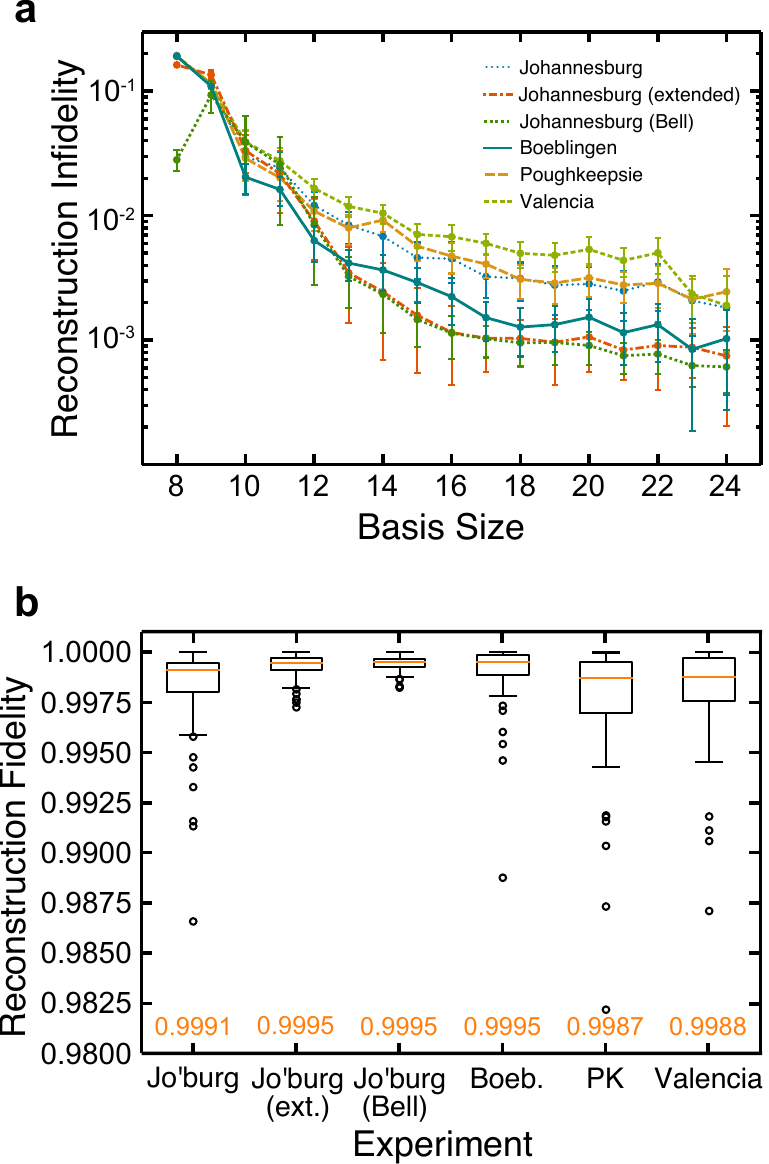}
    \caption{\textbf{Reconstruction fidelity.} For each basis size, we compare the process tensor predictions with experimentally reconstructed density matrices for predictions that lay outside the basis set. \textbf{a} The average infidelity in reconstruction between the states predicted by the process tensor and the experimentally measured state.
    This includes a 95\% confidence interval, computed using the bootstrapping method described in~\cite{bootstrap-method}. The experiments compare the predictions of a basis $n$ process tensor with the experimental outcomes of the $4\times(28-n)\times(28-n)$ experiments from outside the basis set.
    In the notation of the Process Characterisation subsection, our basis is $\mathcal{P}\otimes\mathcal{U}^{(n)}\otimes\mathcal{U}^{(n)}$. \textbf{b} The distribution of fidelities of the predictions made by a basis-24 process tensor over a range of experiments. The top and bottom of the boxes are respectively the 25th and 75th percentiles, the whiskers are 1.5 times the inter-quartile range, and the orange lines are the medians of the distribution, with this last figure also provided in orange to four decimal places.}
    \label{PT-fidelity-summary}
\end{figure}
In theory, a minimal complete basis ($n=10$) is all that is required for a restricted process tensor.
In practice, however, we find that sampling error and, to a lesser extent, gate error, introduces inconsistencies in the linear equations described in Equation~\eqref{DM-expansion}, amplifying reconstruction errors.
The Moore-Penrose pseudoinverse (discussed in the methods section) finds the coefficients minimising the least-squares error between overdetermined and inconsistent linear equations.
Consequently, adding in new basis elements will suppress the noise in the fidelities of prediction.
We find a surprisingly large improvement.
To further minimise bias in the noise, we also order our basis from least to most overlap with the rest of the set, as determined by the Hilbert-Schmidt inner product. 
This basis re-ordering improved predictive fidelity by 20\%.\par 
We summarise the average reconstruction fidelity between prediction and experiment of each basis in Figure~\ref{PT-fidelity-summary}a.
The `Johannesburg (extended)' experiment refers to process tensor experiments with idle time increased by a factor of 32. 
Meanwhile, `Johannesburg (Bell)' is the result of creating a Bell pair, and then acting the unitaries on one half. The intention of these is to probe different dynamics of the system: the former to add a longer time-scale, and the latter to test an initially correlated state. Standard CP maps cannot describe the reduced dynamics of initially entangled states with the environment~\cite{arXiv:1011.6138, PhysRevLett.114.090402}, and so this evaluates a regime in which the process tensor is in principle more applicable.
The results both demonstrate the effects of basis size on process tensor performance, and showcase its ability to characterise processes.
Adding in new basis elements offers substantial improvement in comparison to a minimal complete basis.
Most of the error in reconstruction is statistical. The effects of this can be observed in the three highest fidelity experiments, `Johannesburg (extended)', `Johannesburg (Bell)', and `Boeblingen'. The first two produce more mixed final states, whose density matrices are naturally closer together, and the third is performed with 4096 shots per circuit, compared with 1600 for the remainder.
For a more fine-grained view, Figure~\ref{PT-fidelity-summary}b shows box plots of the predictive fidelity distribution of a size-24 basis on each experiment.
At this size, the median fidelity of characterisation is well within shot noise.
Here, we have shown how to extract useful and accurate predictions, and how unbiased and overcomplete basis sets are necessary for complete practical determination of the process tensor. 
\subsection*{Bounding memory and comparison with GST}
\label{sec:memory-detection}
The impetus of the previous section was to demonstrate an experimentally verifiable method of characterising arbitrary dynamics.
We now show that the above processes are indeed non-Markovian by lower bounding the memory in QIPs. 
We will then show that process tensors make more accurate predictions than comparable Markov models constructed using gate set tomography.
\label{ssec:memory-size}
To fully account for the non-Markovianity in a system requires in-situ measurements, which break all correlations between the system and its environment, and represent a clean barrier to any past-future dependence~\cite{Pollock2018}.
Barring access to these, a restricted process tensor can only infer aspects of the non-Markovianity.
Here, we introduce one such method to extract a lower bound on non-Markovianity.
\par
\begin{figure}
    \centering
    \includegraphics[width=0.95\linewidth]{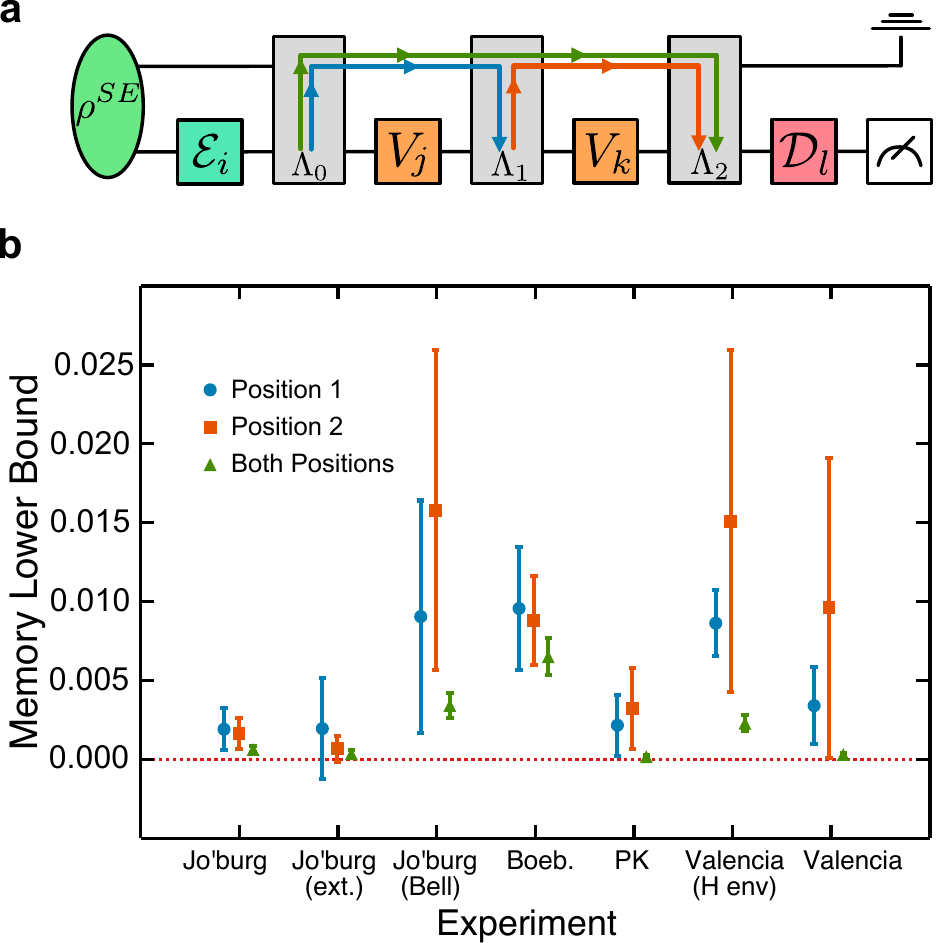}
    \caption{\textbf{Memory size and structure.} \textbf{a} The circuit depicting the process tensor. Quantum information can travel in and out of the system across one or many operations. Each gate is a place-holder for a larger set. Each $V_i$ is an arbitrary unitary operation that need not belong to the set $\mathcal{U}$. \textbf{b} The maximum CMI, which is a conservative lower bound for non-Markovian memory, through $\mathcal{R}$ for each process tensor experiment, with 95\% confidence intervals. This shows statistically significant non-zero memory in the device, which shows consistency in the timescale and the environmental interactions present.}
    \label{fig:mutual_information_circuits}
\end{figure}

Because the maximally depolarising channel
\begin{gather}
\mathcal{R}[\rho] = \frac{\mathbb{I}}{d},\:\forall \: \rho
\end{gather}
lies within the span of unitary operations, we can use it as an information barrier between time steps.
A non-zero mutual information between the input operation and final measurement suggests information has travelled into the environment and returned after $\mathcal{R}$ has been applied~\cite{taranto3}.
Figure~\ref{fig:mutual_information_circuits}a illustrates this idea for the processes we consider here, where $\mathcal{R}$ takes either the first operation position, the second, or both.
This tests the timing and duration of different memory effects.
\par 
The utility of the process tensor here is that it enables us to numerically search for the encoding and decoding operations which give the largest lower bound to non-Markovianity along different paths.
Respectively, these are sets $\mathcal{E}$ and $\mathcal{D}$, the first of which contains two unitary operations applied with equal likelihood, and the second contains two orthogonal measurement effects.
The quantities we compute are the conditional mutual information (CMI) for each case:
\begin{align}
\label{MI-1}
    &\argmax_{\mathcal{E},V_1,\mathcal{D}} I(E:D|\mathcal{E},V_1,\mathcal{R},\mathcal{D}),\\
\label{MI-2}
   & \argmax_{\mathcal{E},V_2,\mathcal{D}} I(E:D|\mathcal{E},\mathcal{R},V_2,\mathcal{D}),\\
\label{MI-3}
    &\argmax_{\mathcal{E},\mathcal{D}} I(E:D|\mathcal{E},\mathcal{R},\mathcal{R},\mathcal{D}),
\end{align}
where:
\begin{gather}
I(E:D) = \sum_{e\in\mathcal{E}} \sum_{d\in\mathcal{D}} p_{(E,D)}(e,d)\log \left(\frac{p_{(E,D)}(e,d)}{p_E(e)p_D(d)}\right).    
\end{gather}
For each experiment, we summarise the memory lower bound in Figure~\ref{fig:mutual_information_circuits}b. 
Note that we include an extra experiment `Valencia (H env)', in which the neighbouring qubits are initialised into the $\ket{+}$ state.
In almost every case, we find non-zero CMI, flagging non-Markovianity within the device.
The extended Johannesburg experiment is the only case for which CMI overlaps zero in all three tests. Given that the effects are no longer observable on this longer timescale, this suggests that the memory has a finite lifetime which can be loosely upper-bounded by this experiment. This is further shown with the lower values where $\mathcal{R}$ is contracted in both positions.
The memory size is especially high for the experiments with coherent neighbours (`Joburg (Bell)' and `Valencia (H env)'), suggesting a passive crosstalk interaction might account for some of the environmental memory effects observed.
\label{ssec:coherent}
The results of Figure~\ref{fig:mutual_information_circuits}b suggest a coupling between neighbouring qubits on Johannesburg and Valencia (we did not assess whether the same effect was present on Boeblingen or Poughkeepsie).
These dynamics provide a useful test-bed for the performance of the process tensor in a non-Markovian system when compared to a Markovian model for the process. 
GST, introduced in~\cite{gst-2013}, is a comprehensive tomographic procedure for estimating process matrices representing gate operations, preparations and measurements.
The maximum likelihood estimate of a gate set employs a Markov model, where repetitions of the gate are taken to be matrix powers.
\par
\begin{figure}
    \centering
    \includegraphics[width=0.95\linewidth]{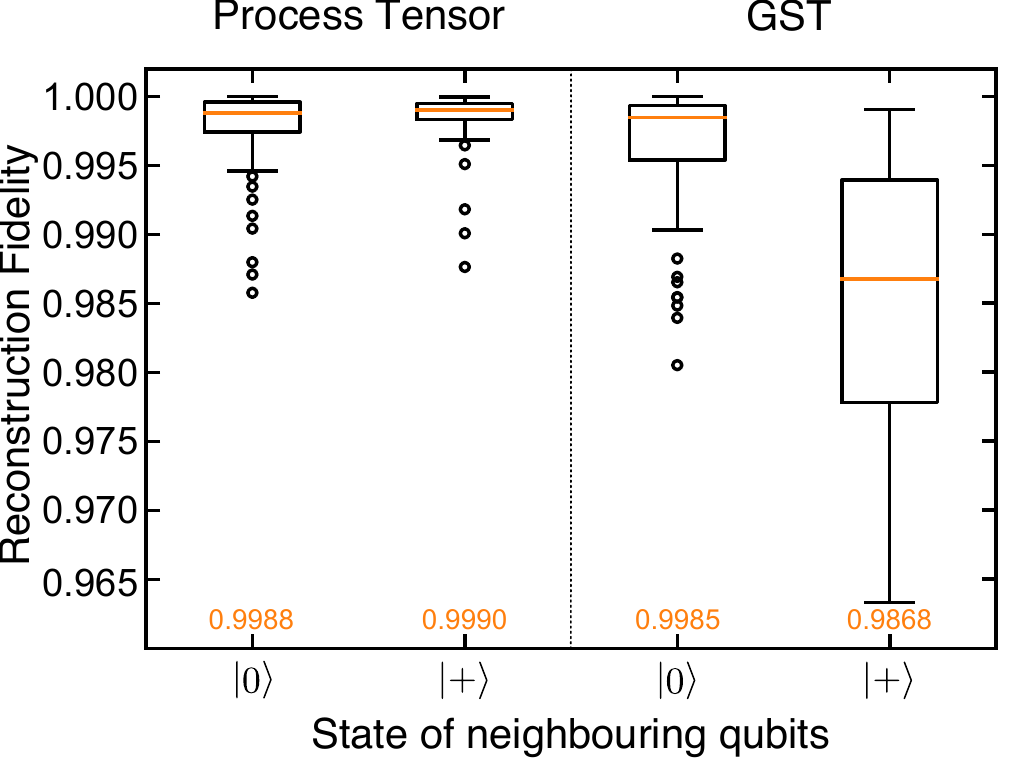}
    \caption{\textbf{Comparison with a Markov model.} We benchmark the accuracy with which different techniques can predict the outcome of a given process for 64 circuits. When nearby qubits are initialised as $\ket{0}$, the median fidelity from GST is similar to the process tensor in each scenario. When the neighbouring qubits are in state $\ket{+}$, however, GST suffers from a fidelity drop of about $1.2\%$. This is a demonstration of how a technique like the process tensor could complement existing characterisation techniques in realistic non-Markovian settings.}
    \label{fig:valencia_boxplots}
\end{figure}

We performed two experiments under two different scenarios on the ibmq\_valencia 5-qubit quantum device. 
The first is identical to the process tensor experiments the Experimental Implementation subsection using the set $\mathcal{U}$.
In addition, using GST we characterised all 28 unitary operations in $\mathcal{U}$, the 4 preparations in $\mathcal{P}$, as well as the the initial state and the final measurement. 
The estimates for each map were multiplied out to produce a Markovian prediction for the final density matrix. 
Both the process tensor and GST experiments were conducted first with neighbouring qubits initialised in the $\ket{0}$ state, and then again initialised in the $\ket{+}$ state.
Figure~\ref{fig:valencia_boxplots} shows the distribution of the reconstruction fidelities for both the process tensor and GST.
With a coherent environment, GST performs about $1.2\%$ worse.
The process tensor tends to perform better in cases where the final state density matrices are more mixed, because this necessarily suppresses any directional bias in the noise.\par
We emphasise that our comparison of the outcomes of the two techniques is not framed competitively.
Indeed, they are qualitatively different: while GST estimates the stationary maps of a given (presumed composable) gateset, the process tensor characterises all possible outcomes in a set process.
Figure~\ref{fig:valencia_boxplots} observes the breakdown of a Markov model, and benchmarks the process tensor against the state-of-the-art as a complementary tool to describing processes. 
\subsection*{Control in the presence of memory}
\label{sec:applications}
In addition to non-Markovian characterisation and diagnostics, we now show that the process tensor can be a useful tool for quantum control.
With a direct map from control operations to experimental outcomes, the data can be used to find which gates optimally output a desired state in a parametrised circuit. 
This outcome could harness external couplings to that end, using only local operations to manipulate them. 
Having already captured the process, the need for hybrid quantum-classical optimisation is eliminated. 
The desired result could be the most entangled state, highest fidelity equal superposition, or some member of a decoherence-free subspace.
The procedure naturally accounts for any mitigating background, such as environmental noise or crosstalk. 
It is a matter of simple numerical optimisation to find the sequence of operations achieving the closest possible state to the one we desire:
(i) Select an objective function $\mathcal{L}$ which computes some quantity on the output density matrix, subject to the sequence $\textbf{A}_{k-1:0}$ of operations performed.
(ii) Find:
\begin{gather}
\label{best-op}
    \argmin_{\textbf{A}_{k-1:0}} \mathcal{L}\left(\mathcal{T}^{k:0}[\textbf{A}_{k-1:0}]\right).
\end{gather}
For unitaries, this is a straightforward minimisation over three parameters per time-step.\par
\begin{figure}
    \centering
    \includegraphics[width=0.85\linewidth]{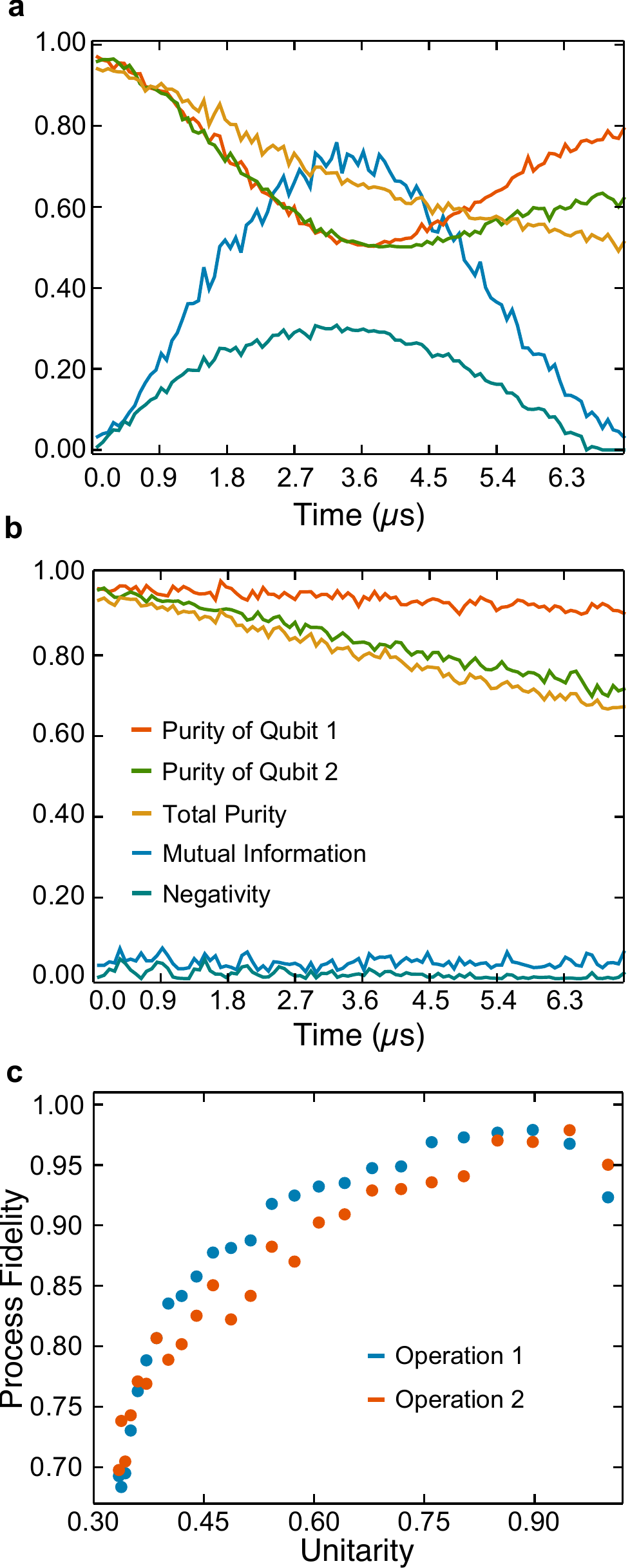}
    \caption{\textbf{Coherent control with non-Markovian noise.} Entanglement, mutual information, and purities extracted from the two-qubit density matrix after being initialised in the $\ket{++}$ state. (a) Both qubits are left idle and the natural evolution is tracked. (b) As a simple demonstrative application of the process tensor, we use the construction from Equation~\eqref{best-op} to find the optimal decoupling pulse. We periodically apply this gate to qubit 1. We see greatly improved coherences and almost complete elimination of entanglement between the two qubits, without actually characterising the nature of the interaction. (c) We use the process tensor to implement specific non-unitary gates. We plot the process fidelity as a function of the unitarity for two randomly chosen operations, according to the measure given in~\cite{noise-coherence-2015}.}
    \label{coherence-plots}
\end{figure}

As an example, we first consider two neighbouring qubits initialised in the $\ket{+}$ state. 
Figure~\ref{coherence-plots}a shows the consequences of their natural coupling, extracted from the reconstructed two-qubit density matrix after some idle time.
The results, which summarise negativity, mutual information, and state purities, show genuine entanglement between the two qubits.
This form of dynamical behaviour will give rise to correlated errors in devices.
After detection of a non-trivial interaction, we can use Equation~\eqref{best-op} to decouple the qubits.
So-called `bang-bang' decoupling approaches have been thoroughly studied in the literature, but usually require \emph{a priori} knowledge of the system-environment interaction Hamiltonian~\cite{qec-textbook}.
Using a one-step process tensor to form outcomes, our objective function is $2-\gamma_1-\gamma_2$, where $\gamma_i$ is the purity of the reduced state: $\gamma_i = \text{tr}(\rho_i^2)$.
Performing the minimisation in Equation~\eqref{best-op}, we find the best decoupling operation. This turns out to be the gate 
\begin{gather}
    \begin{pmatrix}
    0.0051 & \text{e}^{-i\cdot(1.073)} \\
    \text{e}^{i\cdot(0.188)} & 0.0051\cdot\text{e}^{i\cdot(2.257)}
    \end{pmatrix}
\end{gather} 
which amounts to a rotation of approximately $\pi$ around the axis $(n_x,n_y,n_z) = (0.8076,0.5894,4.609\times 10^{-3})$
We then repeat the experiment of Figure~\ref{coherence-plots}a, but periodically apply the decoupling operation approximately every 0.5 $\mu s$.
This yields the results in Figure~\ref{coherence-plots}b, wherein the purities of each qubit have been significantly increased, and the entanglement over time suppressed. Note that this is a demonstration of how the process tensor can be applied as an outcome-based control tool, rather than a rigorous benchmark of decoupling. We have not compared this to standard decoupling techniques, and the operation spacing times were arbitrarily chosen. For further information, see the methods section.
\par
We apply this same technique to exploit non-Markovianity for enhanced quantum control, inspired by the ideas in~\cite{non-Markovian-control}.
Arguments for the use of non-Markovianity as a resource are founded upon accessing Hilbert space trajectories otherwise unavailable with system control.
We broaden our control set by using the process tensor to include non-unitarity, limited only by the strength and duration of the underlying interaction.
We achieve this by constructing a single-step process tensor on one half of a pair of coupled qubits for a set of four preparation operations.
Then, we use Equation~\eqref{best-op} to find the parameters that produce final states closest to the ideal outputs of a randomly selected non-unitary operation,
before applying the corresponding gate and performing quantum process tomography on it.
The process fidelity of these non-unitary maps compared to their targets is plotted as a function of unitarity in Figure~\ref{coherence-plots}c.
It reaches up to $97\%$, showing that we can extend the control capabilities of the device by using the process tensor and a nearby coupled qubit. This target gate is achieved for a given interaction of the system with its neighbour. Since interaction time is not varied, the maximum achievable non-unitarity is fixed, which is why the process fidelity decreases when gates with a lower unitarity are targeted.
This shows a way forward in which extended control regimes could be used for the implementation of non-unital and trace-decreasing maps which are necessary for the reconstruction of the full process tensor.
Critically, for this to work, we do not need to perform control operations on the neighbouring qubit beyond its initialisation. For further detail about this implementation, see the methods section.\par 
This simple framework is widely applicable to many forms of quantum control. 
In particular, it allows for either mitigating or controlling non-Markovian noise without first understanding it at a microscopic level. 
Broadly, the user need only specify a desired outcome, without studying the means to achieve it.
\section*{Discussion}
In this paper, we have bridged the gap from a theoretical framework of non-Markovian dynamics to an experimental method which verifiably offers non-Markovian diagnostics and control. First, we demonstrated a high fidelity non-Markovian characterisation technique over a range of devices. 
We used this to bound the non-Markovian memory present.
Then, using the reconstructed process tensor, we demonstrated operationally tractable control techniques to decouple the system qubit from its neighbour, as well as applying well-characterised intermediary non-unitary operations on the system. These methods pave the way to mitigate non-Markovian noise and streamline the performance of quantum devices. Although tested on superconducting qubits, the principles behind this technique are agnostic to the hardware. Implementation of the control operations across different platforms would be a useful avenue to explore in future work.\par
Like many tomographic techniques, the construction of the process tensor scales unfavourably in both the number of time-steps and number of qubits. 
However, for processes with finite Markov order it is possible to reconstruct a primitive building block, from which the whole process can be inferred~\cite{taranto3}.
One immediate future avenue is complete process characterisation, as suggested in the previous section, which will offer better benchmark for the length of the memory. 
Although we found success with the use of an overcomplete basis, it would likely be fruitful to explore coupling smaller bases with conventional denoising techniques, the use of a mutually unbiased unitary basis~\cite{Nasir2020}, or machine learning reconstruction methods~\cite{guochu}.
Much like with the study of many-body entanglement, there is ample room to reduce experimental overhead with some well-placed physical assumptions.

\section*{Methods}
\label{sec:methods}
\subsection*{Process Tensor Experiments}
\begin{figure}
    \centering
    \includegraphics[width=\linewidth]{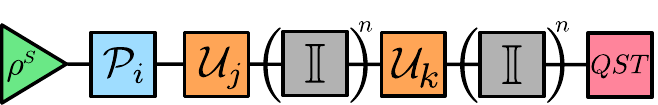}
    \caption{Circuit diagram depicting the generic experiment required to construct the two step process tensor. Each gate represents an element from either the preparation set $\mathcal{P}$ or the more general unitary basis set $\mathcal{U}$. The identity gates represent idle time which we allow to vary. Finally, measurements in three bases are made for QST.}
    \label{methods-PT}
\end{figure}

Here, we discuss the construction of a multi-time process tensor both in particular to the experiments conducted in this work, and more generally with respect to a greater set of controls.
The process tensor constructed was over three time-steps of varying sizes. The experimental steps for this are as follows:
\begin{enumerate}
    \item Initialise the qubit in state $\ket{0}$
    \item Apply $\mathcal{P}^i \in \mathcal{P} = \{H,S\cdot H, \mathbb{I},X\}$
    \item Apply $\mathcal{U}^j \in \mathcal{U}$
    \item Leave  some amount of time.
    \item Apply $\mathcal{U}^k \in \mathcal{U}$.
    \item Leave idle.
    \item Repeat this sequence three times for the three QST basis measurements required.
    \item Store this density matrix as $\rho_3^{ijk}$
    \item Repeat this for all combinations of the elements of $\mathcal{P}$ and $\mathcal{U}$ in each slot.
\end{enumerate}
For our experiments, this is a total of $(4\times 28\times 28)\times 3 = 9408$ experiments.
Interleaved between each operation is idle time equivalent to a single gate. The circuit diagram for these experiments is given in Figure~\ref{methods-PT}. 
We ran these at 1600 shots each with the exception of `Boeblingen', which had 4096 shots. 
This data was then partitioned into process tensor construction, and experimental verification.
The former consists of the construction of a basis-$n$ process tensor, which used the first $4\times n \times n$ control sequences to form a basis.
We then used the remaining $4\times (28-n) \times (28-n)$ sequences which lie outside the basis set as verification density matrices for the process tensor predictions.
It is worth noting that action of the process tensor is insensitive to state preparation and measurement (SPAM) errors.
Any initial state or final measurement error channel are absorbed into the definition of the process tensor, and the expansion remains the same.\par 
The unitary basis was constructed with a randomly generated set of 28 ordered unitary matrices using the \texttt{scipy.stats.unitary\_group.rvs()} function.
We parametrise these gates using the standard \texttt{qiskit} unitary parametrisation: 
\begin{gather}
\label{unitary-param}
    U(\theta,\phi,\lambda) = \begin{pmatrix}
\cos(\theta/2) & -e^{i\lambda}\sin(\theta/2) \\
e^{i\phi}\sin(\theta/2) & e^{i\lambda+i\phi}\cos(\theta/2) 
\end{pmatrix}.
\end{gather}
On the IBM superconducting devices, these so-called \texttt{u3} gates are implemented in two physical pulses corresponding to rotations around the $x-$axis, and three frame shifts corresponding to rotations around the $z-$axis~\cite{qasm-doc,PhysRevA.96.022330}. 
Explicitly, 
\begin{gather}
    U(\theta,\phi,\lambda) = R_z(\phi + 3\pi)R_x(\pi/2)R_z(\theta+\pi)R_x(\pi/2)R_z(\lambda).
\end{gather}
Consequently, the physical duration of each \texttt{u3} gate is independent of the $\theta,\phi,\lambda$ parameters -- approximately 72 ns.
We then leave the system idle for a duration of one \texttt{u3} gate.
Following this is one more \texttt{u3} gate, an identical wait time, and then each of three basis measurements in $X$, $Y$, and $Z$ Pauli bases required to reconstruct the output density matrix.
The maximum likelihood method introduced in~\cite{PhysRevLett.108.070502} is then used to find the closest physical density matrix consistent with the data.
The ordered list of density matrices collected make up the experimental data required for the process tensor.\par 
The IBM Quantum devices are fixed-frequency superconducting transmon devices, each with similar error rates and coherence times. ibmq\_boeblingen, ibmq\_poughkeepsie, and ibmq\_johannesburg are each 20 qubits, while ibmq\_valencia is a five qubit processor. 
\subsection*{Control basis and process reconstruction}
An arbitrary $\mathcal{A}_{j}$, at time step $j$, on a system of dimension $d_S$ may be decomposed into a linear expansion of some ordered basis $\{\mathcal{B}_j^{\mu_j}\}$ such that
\begin{gather}
    \mathcal{A}_j = \sum_{\mu_j=1}^{d_S^4} \alpha^{\mu_j}_j \mathcal{B}^{\mu_j}_j.
\end{gather}
A sequence of (independently chosen) control operations may be written with a tensor product structure $\mathbf{A}_{k-1:0} = \bigotimes_{j=0}^{k-1} \mathcal{A}_{j}$, for which each constituent map can be further decomposed into the chosen basis. The complete spatio-temporal basis of operations is then given by 
\begin{gather}
\left\{\mathbf{B}_{k-1:0}^{\bm{\mu}} = \bigotimes_{j=0}^{k-1} \mathcal{B}_j^{\mu_j} \right\}_{\bm{\mu}=(1,1,\cdots,1)}^{(d_S^4,d_S^4,\cdots,d_S^4)}    ,
\end{gather}
where $\bm{\mu} = (\mu_0, \mu_1, \cdots, \mu_{k-1})$ is a $k-$dimensional vector of index elements, each taking values between $1$ and $d_S^4$. That is, it is the set with cardinality $d_S^{4k}$ of all combinations of the $k$ tensor products of each member of $\{\mathcal{B}^{\mu_j}_j\}$ at each time step. 
Measuring the output state $\rho_k^{\bm{\mu}}$ for each of these basis operations is sufficient to construct the process tensor. We signify the matrix form of the process tensor $\mathcal{T}$ with a caret: $\hat{\mathcal{T}}$
\begin{gather}
    \hat{\mathcal{T}}^{k:0} = \sum_{\bm{\mu}} (\mathbf{\Delta}^{\bm{\mu}}_{k-1:0})^\text{T} \otimes\rho_k^{\bm{\mu}},
\end{gather}
where the set $\{\mathbf{\Delta}_{k-1:0}^{\bm{\mu}} 
\}$ 
is known as the dual set to $\{\hat{\mathbf{B}}_{k-1:0}^{\bm{\mu}}\}$, satisfying $\text{tr} \left[\hat{\mathbf{B}}_{k-1:0}^{\bm{\mu}} \mathbf{\Delta}_{k-1:0}^{\bm{\nu}} \right] = \delta_{\bm{\mu} \bm{\nu}}$. This dual set can be easily computed for any linearly independent set of vectors.
To be explicit, the matrix form for the two-step process tensor using a basis of $n$ operations is given as 
\begin{gather}
\label{two-step-pt}
    \hat{\mathcal{T}}^{3:0} = \sum_{i=1}^4\sum_{j=1}^n\sum_{k=1}^n  \left(D^i_0\otimes \Delta^j_1 \otimes\Delta^k_2 \right)^{\text{T}}\otimes\rho^{ijk}_3
\end{gather}
Where the $\{D^i_0\}$ are dual to the preparation operations $\mathcal{P}$, and the $\{\Delta^j_1\} = \{\Delta^k_2\}$ are dual to the circuit operations $\mathcal{U}^{(n)}$.
Sampling error in the final state density matrix, as well as error in the gates themselves will collectively introduce inconsistencies in the set of linear equations described by Equation~\eqref{DM-expansion}.
The error becomes significant if the basis is biased in a particular direction of superoperator space.
Originally, our minimal complete basis -- which had been randomly selected -- produced a reconstruction fidelity of around $70\%$.
To mitigate this error, we re-ordered our basis according to the least to most overlap with the remainder of the set according to the Hilbert-Schmidt inner product.
For the first ten elements, this overlap was $[0.0336,0.0409,0.0438,0.0489,$ $0.0505,0.0518,0.0594,0.0600,0.0619,0.0621]$. After re-ordering, the reconstruction fidelity of the minimal complete basis improved to around $95\%$.
This effect was only discovered after the completion of the experiments, at which point it was too late to change the operational basis itself.
In future, a better course of action would be to examine the selection of a set of mutually unbiased unitary operators.\par
In general, the only positive dual operators are entanglement-breaking channels. 
With a restricted basis, the process tensor constructed here is not unit trace, nor is it a positive operator.
Physically meaningful quantities can only be extracted from its action on the restricted basis, rather than from the explicit form given in Equation~\eqref{two-step-pt}.
For this reason, we keep the emphasis of the process tensor in this work on its `actions' rather than on information that can be gleaned from the object itself. 
Note that the expansion coefficients are calculated in the contraction of the operation with the process tensor.
We discuss this explicitly below. 
\subsection*{Construction of a dual set}

The procedure to construct the dual operators is as follows: for a complete set of linearly independent operations $\{\mathcal{B}^i\}$ whose matrix forms are $\{\hat{\mathcal{B}}^i\}$, we can compile the basis into a single matrix $\mathscr{B}$.
Write each $\hat{\mathcal{B}}^i = \sum_j b_{ij}\Gamma_j$, where $\{\Gamma_j\}$ form a Hermitian, self-dual, linearly-independent basis satisfying $\text{tr}[\Gamma_j\Gamma_k]=\delta_{jk}$. 
In our case, we select $\{\Gamma_j\}$ to be the standard basis, meaning that the $k$th column of the matrix $\mathscr{B} = \sum_{ij}b_{ij}\ket{i}\bra{j}$ is $\hat{\mathcal{B}}^k$ flattened into a $1$D vector. 
Because the $\{\hat{\mathcal{B}}^i\}$ are linearly independent, $\mathscr{B}$ is invertible. 
Let the matrix $\mathscr{F}^\dagger = \mathscr{B}^{-1}$ such that $\mathscr{B}\cdot \mathscr{F}^\dagger = \mathbb{I}$.
This means that the rows of $\mathscr{F}^\dagger$ are orthogonal to the rows of $\mathscr{B}$.
The dual matrices can then be defined as $\Delta^i = \sum_j f_{ij}\Gamma_j$, ensuring that $\text{tr}[\hat{\mathcal{B}}^i\Delta^j] = \delta_{ij}$.
Note that in this work, our basis is restricted to the sub-manifold of unitary matrices.
This means that the dimension $d$ of the space is less than the order $n$ of the matrices. 
Therefore we construct $\mathscr{F}^\dagger$ as the Moore-Penrose or the right inverse of $\mathscr{B}$.
We also primarily operate in an over-complete setting, where the number of basis operations is greater than the dimension of the space, meaning that they cannot all be linearly independent.
Here, we relax the duality condition $\text{tr}[\hat{\mathcal{B}}^i\Delta^j]=\delta_{ij}$, but retain $\sum_i\Delta^i = \mathbb{I}$ to ensure that the expansion of any operation within the basis is complete. 
The over-completeness technique is necessary for a high fidelity reconstruction, owing to the sensitivity of the matrix pseudoinverse to shot-noise.\par 
\subsection*{Contracting an Operation}
The expansion coefficients discussed are useful in conceptual discussions of the process tensor, but in practice these are not directly computed.
Instead, the action of the process tensor on a sequence of operations is found by projecting the process tensor onto the Choi state of this sequence (up to a transpose).
Below, we explicitly step through this computation. 
\begin{gather}
\begin{split}
        &\mathcal{T}^{k:0}\left[\mathbf{A}_{k-1:0}\right] = \text{tr}_{\text{in}} \left[\left(\hat{\mathbf{A}}_{k-1:0} \otimes\mathbb{I}_{\text{out}}\right)^\text{T}
        \hat{\mathcal{T}}^{k:0}\right]\\
        &=\text{tr}_{\text{in}}\left[\left(\bigotimes_{i=0}^{k-1}\hat{\mathcal{A}}_i^\text{T} \otimes \mathbb{I}\right)\sum_{\bm{\nu}} (\mathbf{\Delta}^{\bm{\nu}}_{k-1:0})^\text{T}\otimes \rho_k^{\bm{\nu}}\right]\\
        &=\text{tr}_{\text{in}}\left[
        \sum_{\bm{\mu}}
        \alpha^{\bm{\mu}}
        \bigotimes_{i=0}^{k-1} 
        \hat{\mathcal{B}}_i^{\mu_i \text{T}}
        \sum_{\bm{\nu}} 
        \bigotimes_{j=0}^{k-1}
        \Delta_j^{\nu_j \text{T}} \otimes \rho_k^{\bm{\nu}}\right]\\
        &=\text{tr}_{\text{in}}\left[
        \sum_{\bm{\mu},\bm{\nu}} 
        \alpha^{\bm{\mu}}
        \bigotimes_{i,j=0}^{k-1} \{\hat{\mathcal{B}}_i^{\mu_i \text{T}} \Delta_j^{\nu_j \text{T}}\}\otimes \rho_k^{\bm{\nu}}\right]\\
         &=\sum_{\bm{\mu},\bm{\nu}} 
         \alpha^{\bm{\mu}}
         \prod_{i,j=0}^{k-1}  \,
        \text{tr}\left[
        \hat{\mathcal{B}}_i^{\mu_i} \Delta_j^{\nu_j}\right]
        \rho_k^{\bm{\nu}}\\
        &=\sum_{\bm{\mu},\bm{\nu}}
        \alpha^{\bm{\mu}}
        \prod_{i=0}^{k-1}  \, \delta_{\bm{\mu}\bm{\nu}} \, \rho_k^{\bm{\nu}} \\
        &= \sum_{\bm{\mu}} \alpha^{\bm{\mu}}\rho^{\bm{\mu}}_k\\
        &=\rho_k(\textbf{A}_{k-1:0}).
\end{split}
\end{gather}
The direct calculation of each expansion coefficient is therefore given by \begin{align}
    \alpha^{\bm{\mu}} =& \text{tr}\left[\hat{\mathbf{A}}_{k-1:0}\mathbf{\Delta}_{k-1:0}^{\bm{\mu}}\right]\\
    =& \text{tr}\left[
    \bigotimes_{i=0}^{k-1} \hat{\mathcal{A}}_{i}
    \Delta^{(\mu,i)}\right]\\
    =& \prod_{i=0}^{k-1}
    \text{tr}\left[
     \hat{\mathcal{A}}_{i}
    \Delta_i^{\mu_i }
    \right] = \prod_{i=0}^{k-1} \alpha_i^{\mu_i}.
\end{align}
\subsection*{Bounding Memory}
In the Bounding Memory subsection, we estimate a lower bound for the memory present in the devices.
This is accomplished with the contraction of different encoding operations with the process tensor and forming predictions for the output in this way. For the case where $\mathcal{R}$ is contracted in position one, the explicit steps are as follows:
\begin{enumerate}
    \item Pick $\mathcal{E}^0, \mathcal{E}^1 \in U(2)$
\item Pick $p_{e_0}$ and $p_{e_1}$ s.t. $p_{a_i}\in [0,1]$ and $p_{e_0} + p_{e_1} = 1$ (in this experiment, we set $p_{e_0} = p_{e_1} = 0.5$).
\item Pick $\mathcal{D} \in U(2)$ 
\item Pick $\mathcal{V} \in U(2)$ 
\item Compute the 4 values of $p_{(E,D)}(e_i,d_j)$ by collecting the density matrix $\rho^i = \mathcal{T}^{3:0}[\mathcal{E}^i, \mathcal{V},\mathcal{R}]$ and then setting $p_{(E,D)}(e_i,d_j) = p_{e_i}\cdot \text{Tr}(|j\rangle \langle j| \cdot \mathcal{D}\rho_i \mathcal{D}^\dagger)$
\item Compute the marginal distributions: $p_E(e_i) = \sum_j p_{(E,D)}(e_i,d_j)$ and $p_D(d_j) = \sum_i p_{(E,D)}(e_i,d_j)$
\item Finally, compute $I(E : D)$
\end{enumerate}
These steps are framed as an optimisation problem where $\mathcal{E},\mathcal{D},$ and $\mathcal{V}$ are chosen such that $I(E:D)$ is maximised.
\begin{figure}
    \centering
    \includegraphics[width=\linewidth]{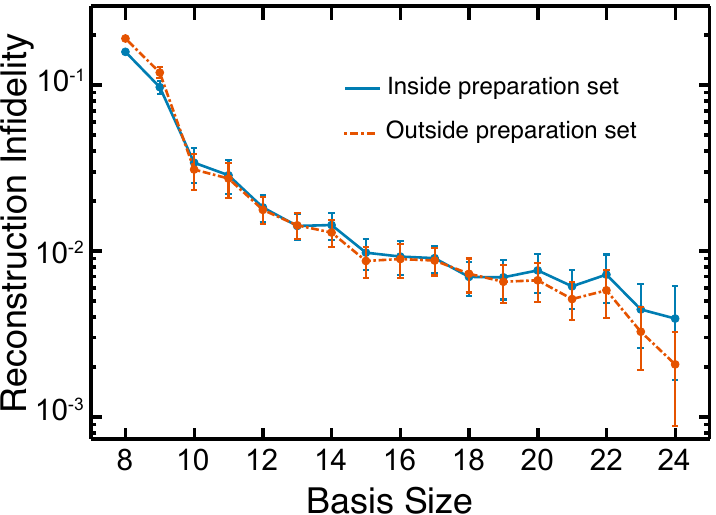}
    \caption{\textbf{Reconstruction infidelity for a process tensor experiment on the ibmq\_valencia.} Here, we examine a four time process tensor whose basis is $\mathcal{U}^{(4)}\otimes \mathcal{U}^{(n)}\otimes\mathcal{U}^{(n)}$. We compare the reconstruction fidelity between predictions made for the experimental sequences $\mathcal{U}^{(1:4)}\otimes\mathcal{U}^{(n:28)}\otimes\mathcal{U}^{(n:28)}$ (inside the preparation set) with $\mathcal{U}^{(5:8)}\otimes\mathcal{U}^{(n:28)}\otimes\mathcal{U}^{(n:28)}$ (outside the preparation set). We find that they are, within error, the same. Indeed with slightly better performing results for the unitaries outside of the basis set.}
    \label{methods-memory}
\end{figure}
Implicit in this exercise is the assumption that operations outside the preparation set achieve the same reconstruction fidelity as the latter steps shown in Figure \ref{PT-fidelity-summary}. 
Although we did not examine this assumption for every machine, in Figure \ref{methods-memory} we construct a four time process tensor on Valencia using the basis $\mathcal{U}^{(4)}\otimes \mathcal{U}^{(n)}\otimes \mathcal{U}^{(n)}$. 
We then compare the reconstruction infidelity from predictions made by the process tensor: firstly, compared to gate sequences where the preparation operation lay inside the basis set (with $\mathcal{U}_1^i$ and $\mathcal{U}_2^j$ outside), and secondly compared to gate sequences where the preparation operations were the next four elements of $\mathcal{U}$.
We find these two collections to be identical within error bars for all basis sizes 10 and above. 
Given that Valencia had the worst reconstruction fidelity of the machines, we view this as sufficient evidence that the assumption is valid across all machines.

\subsection*{Gate set tomography comparison}
The GST experiments conducted in the Comparison with GST subsection were completed using the \texttt{pyGSTi} quantum processor performance package~\cite{pygsti,pygsti-arxiv}.
Following the procedures outlined in the documentation, with background given in~\cite{intro-GST,sandia-GST}, we characterised the 28 random unitaries as well as the 4 preparation gates in 8 groups of 4 gates. 
The software package designates the circuits required, and carries out the maximum likelihood reconstruction of the gates with the constraint of complete positivity and trace preservation. 
The gate sequences were repeated in powers of 2: $1, 2, 4, 8, 16,$ and $32$ times.
Included in this estimate are the state preparation and measurement vectors, $\kket{\rho}$ and $\bbra{E}$.
The process tensor and GST experiments were conducted in the one calibration period for the device in a window of approximately 5 hours.
The gates were characterised in different groups for computational convenience, however this means that the final estimate for each group cannot necessarily be mixed. 
There exists a gauge freedom in gate set tomography, in which measurement outcomes $\bbra{E} G\kket{\rho}$ is invariant under the transformation $\bbra{E} \mapsto \bbra{E}B$, $\kket{\rho} \mapsto B^{-1}\kket{\rho}$ and $G\mapsto B^{-1}GB$. In the GST estimate, this gauge is optimised to bring the gate set as close as possible to the target set. However, in principle, each of the sets characterised will be in slightly different gauges. In order to estimate the effects of this, we computed the reconstruction fidelities with respect to the SPAM vector estimates of each gate set estimate. Of the different gauges, the one with the maximum average difference between the data points in any of these distributions and the for the $\ket{+}$ neighbour given in Figure \ref{fig:valencia_boxplots} is $5.9\times 10^{-3}$, which is similar in magnitude to sampling error and does not significantly affect the comparison. This suggests that the absolute performance of the GST estimates could be marginally better than what is shown.
\par

\subsection*{Adaptive control methods}
\begin{figure}
    \centering
    \includegraphics[width=\linewidth]{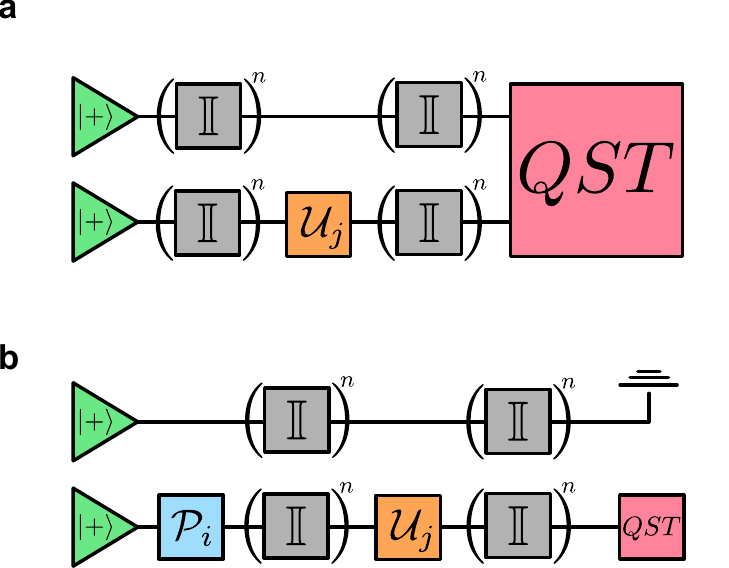}
    \caption{\textbf{Circuit diagrams for each of the application experiments.} \textbf{a} For the decoupling of two qubits, we allow evolution time before and after the process tensor. The distinction between here and other process tensor experiments that we conducted is that we map from the operation on one qubit to the two-qubit density matrix, rather than solely single qubits. \textbf{b} To enact non-unitary gates of our choosing, we conduct a similar experiment. This time, however, there are four basis preparation operations to begin with, and QST only on the single qubit. This is so that we can optimise the action of the gate over a complete basis of inputs.}
    \label{methods-applications}
\end{figure}

Here, we more explicitly discuss our adaptive control methods using the process tensor. 
In each case, the system qubit and its neighbour were both initialised in the $\ket{+}$ state.
We sought to use the process tensor to control the always-on interaction between the two qubits without actually learning it.
The circuit diagrams describing both experiments are in Figure~\ref{methods-applications}.
In the first scenario, using operations only on qubit 1, we construct a single step process tensor with a size-24 basis, 256 ns of idle time on either side, and two-qubit state tomography at the end.
Altogether, this is $24\times 9 = 216$ experiments.
Strictly speaking only single qubit state tomography is required for the purpose of decoupling one qubit, however we created a mapping to the two-qubit output in order to specifically best show these two qubits decoupled.
With the intermediate operation parametrised as in Equation~\eqref{unitary-param}, the minimisation performed was:
\begin{gather}
\begin{split}
    &\argmin_{\theta,\phi,\lambda} 2 - \gamma_1 - \gamma_2
\end{split}
\end{gather}
where $\gamma_i$ is the purity of the $i$th reduced density matrix produced by the process tensor.
The total density matrix is $\mathcal{T}^{1:0}\left[U(\theta,\phi,\lambda)\right]$.
The decoupling operation found was 
\begin{gather}
    \begin{pmatrix}
    0.0051 & \text{e}^{-i\cdot(1.073)} \\
    \text{e}^{i\cdot(0.188)} & 0.0051\cdot\text{e}^{i\cdot(2.257)}
    \end{pmatrix}.\notag
\end{gather}
This amounts to a rotation of approximately $\pi$ around the axis $(n_x,n_y,n_z) = (0.8076,0.5894,4.609\times 10^{-3})$. In Figure~\ref{coherence-plots}b, we periodically apply this operation to the system after the equivalent amount of time in order to decouple the two qubits. 
\par 
For the purpose of implementing our own chosen non-unitary operations, we created a one-step basis-24 process tensor on a single qubit whose neighbour was in the $\ket{+}$ state: approximately $800$ ns of idle time after $\mathcal{P}$ preparations, followed by $\mathcal{U}^{(24)}$, followed by another $800$ ns and then QST.
We then generated a set of random non-unitary operations with unitarity ranging from $1/3$ to $1.0$.
These are denoted by $\mathcal{N}(\alpha,\eta)$, where
\begin{gather}
\begin{split}
\mathcal{N}(\alpha,\eta) = \sqrt{\eta}\mathcal{E}(\alpha) + \sqrt{1-\eta}Y\mathcal{E}(\alpha),\\
\text{and}\quad \mathcal{E}(\alpha) = (R_X(\alpha)R_Y(\alpha)R_Z(\alpha)).
\end{split}
\end{gather}
The two operations shown in Figure~\ref{coherence-plots}c are two different randomly generated values for $\alpha$. The unitarity of the operations is then varied by varying $\eta$ from $0$ to $0.5$ in the above equation.
Using these operations as a target map, we numerically found the gate parameters minimising the trace distance between the target outputs of the non-unitary map and the process tensor predictions for a set of four inputs.
That is, we applied the minimisation:
\begin{gather}
\begin{split}
    \argmin_{\theta,\phi,\lambda} &\frac{1}{2} \left(||\tau_X - \rho_X||_1 + ||\tau_Y - \rho_Y||_1\right. \\ 
    &+ \left.||\tau_Z - \rho_Z||_1 + ||\tau_{I-Z} - \rho_{I-Z}||_1 \right),
\end{split}
\end{gather}
where each $\rho_j$ is the ideal output of $\mathcal{N}(\alpha,\eta)$ acting on the $X, Y, Z, $ and $\mathbb{I}-Z$ eigenvectors, and each $\tau_j$ is the $\mathcal{T}^{2:0}\left[\mathcal{P}_j, U(\theta,\phi,\lambda)\right]$ predicted density matrices. 
Then, using the optimal values of $\theta, \phi, $ and $\lambda$, we performed quantum process tomography and compared the process tensor of our implementation $\mathcal{N}'(\alpha,\eta)$ with the ideal $\mathcal{N}(\alpha,\eta)$.

\section*{Acknowledgments}
This work was supported by the University of Melbourne through the establishment of an IBM Quantum Network Hub at the University. 
GALW is supported by an Australian Government Research Training Program Scholarship. 
CDH is supported through a Laby Foundation grant at The University of Melbourne. 
KM is supported through Australian Research Council Future Fellowship FT160100073.

\section*{Data Availability}
The datasets generated and/or analysed during the current study are available from the corresponding author on reasonable request.

\section*{Code Availability}
The code developed during the current study is available from the corresponding author on reasonable request.

\section*{Author Contributions}
G.W., C.D.H., F.A.P., L.C.L.H., and K.M. contributed to the practical development of the process tensor as a characterisation technique; G.W. designed and conducted all experiments, code, and data analysis with input from the other authors; G.W. and K.M. wrote the manuscript with input from all authors; K.M., C.D.H., and L.C.L.H. supervised the project.
\section*{Competing Interests}
The authors declare no competing financial interests. 
\textbf{Non-financial competing interests:} The authors are supported by the University of Melbourne through the establishment of an IBM Quantum Network Hub at the University.

\end{document}